%% file: QEC_specialized_proc_V0.tex
\documentclass[aps,pra,
			   reprint,
			   a4paper,
			   longbibliography, 
			   floatfix, 
			  ]{revtex4-1}
\usepackage[T1]{fontenc}
\usepackage[latin9]{inputenc}
\usepackage[english]{babel}
\usepackage{graphicx,amsmath,amssymb,color}
\usepackage[normalem]{ulem}
\usepackage{amsfonts}
\usepackage[toc,page]{appendix}
\usepackage[colorlinks=true,linkcolor=blue,urlcolor=blue,citecolor=blue]{hyperref}
\usepackage{latexsym}
\usepackage{amsfonts}
\usepackage{algpseudocode}
\usepackage{amsthm}
\usepackage{mathrsfs}
\usepackage{color,verbatim}
\usepackage{psfrag}

\input{tikzit/tikz_premable} 

\begin{document}

\title{Decoding quantum error correction with Ising model hardware}
\author{Joschka Roffe${}^a$, Stefan Zohren${}^b$, Dominic Horsman${}^c$, Nicholas Chancellor${}^d$}
\email{nicholas.chancellor@gmail.com}
\affiliation{
${}^a$ Department of Physics \& Astronomy, University of Sheffield, Sheffield, S3 7RH, United Kingdom \\
${}^b$ Machine Learning Research Group and Oxford-Man Institute for Quantitative Finance,\\ 
 Department of Engineering Science, University of Oxford, United Kingdom \\
${}^c$ Laboratoire d'Informatique de Grenoble, Universit\'e Grenoble Alpes, 38000 France  \\
 ${}^d$ Joint Quantum Centre (JQC) Durham-Newcastle, Department of Physics, Durham
	University, United Kingdom
}%

\begin{abstract}

Fault tolerant quantum computers will require efficient co-processors for real-time decoding of their adopted quantum error correction protocols. In this work we examine the possibility of using specialised Ising model hardware to perform this decoding task. Examples of Ising model hardware include quantum annealers such as those produced by D-Wave Systems Inc., as well as classical devices such as those produced by Hitatchi and Fujitsu and optical devices known as coherent Ising machines. We use the coherent parity check (CPC) framework to derive an Ising model mapping of the quantum error correction decoding problem for an uncorrelated quantum error model. A specific advantage of our Ising model mapping is that it is compatible with maximum entropy inference techniques which can outperform maximum likelihood decoding in some circumstances. We use numerical calculations within our framework to demonstrate that maximum entropy decoding can not only lead to improved error suppression, but can also shift threshold values for simple codes. As a high value problem for which a small advantage can lead to major gains, we argue that decoding quantum codes is an ideal use case for quantum annealers. In addition, the structure of quantum error correction codes allows application specific integrated circuit (ASIC) annealing hardware to reduce embedding costs, a major bottleneck in quantum annealing. Finally, we also propose a way in which a quantum annealer could be optimally used as part of a hybrid quantum-classical decoding scheme.

\end{abstract}

\maketitle

\section{Introduction and background}

There have been promising advances in practical quantum computing in recent years and a range of prototype devices are currently in development \cite{qed_ibm,Brandl2016,Harty2014}. Quantum computing architectures can be broadly divided into two classes: gate model devices based on discrete operations, and continuous time devices for which the native dynamics of the quantum system is used to solve problems. For gate model quantum computing, the aim is to build a scalable and fault tolerant device capable of universal logic. In principle, this can be achieved through the use of quantum error correction protocols to achieve arbitrary suppression of noise at the logical level \cite{Preskill385,Kitaev1997,Aharonov:1997:FQC:258533.258579,Knill1998}. In contrast, continuous time quantum computers, in particular the highly successful subfield of quantum annealing, can operate without having to be error corrected \cite{brooke99a,johnson11a,denchev16a,lanting14a,boixo16a}. Quantum annealing devices have already been built with thousands of qubits \cite{D-Wave}. Whilst existing quantum annealing devices are not universal, they have been found to be useful for a wide variety of applications, for example in theoretical computer science \cite{chancellor16a}, finance \cite{marzec16a,Orus18a,Venturelli18a}, aerospace \cite{coxson14a}, machine learning \cite{amin16a,Benedetti16a,Benedetti16b,Adachi2015}, mathematics \cite{Li17a,Bian13a}, decoding of communications \cite{chancellor16b}, hydrology \cite{omalley18a}, and computational biology \cite{perdomo-ortiz12a}.

Quantum error correction protocols impose considerable overhead in the design of fault tolerant quantum computers. For example, the surface code will demand around four thousand physical qubits per logical qubit \cite{gidney18,fowler2012surface}. In addition to the qubit overhead, another consideration is how to efficiently interpret the output of a quantum error correction code in real-time. This task, referred to as decoding, is known in to be computationally demanding \cite{Breuckmann17a}, with data rates from some quantum codes expected to be of the order $100$Gbit/sec.  In this present paper we explore the use of quantum annealers and related Ising model devices as specialised co-processors for decoding.

The task of decoding imposes a bottleneck on the successful operation of quantum error correction (especially in situations where non-Clifford operations are to be performed). Achieving even a small improvement in decoding could therefore lead to major in gains in performance of the quantum computer. As a result, the decoding problem is a high-value use case for quantum annealing devices. In addition, this work demonstrates how the two paradigms in quantum computing architecture can be combined in a hybrid setting.

\subsection{Quantum error correction}

In the gate model of quantum computation, universal computation is achieved through the application of discrete operations from a finite set of qubit gates. Because these gates are realised experimentally via the precise manipulation of fragile quantum systems, both control and memory faults are common. As such, quantum error correction protocols play an essential role in the design of any gate model quantum computer.

Adapting existing classical error correction protocols for use on quantum hardware is not straightforward. The No-Cloning theorem prohibits quantum data from being redundantly encoded via simple duplication \cite{Wootters1982}. Furthermore, quantum error correction protocols must be carefully designed to avoid compromising the encoded information via wavefunction collapse. A final complication that arises when designing quantum codes is that qubits are susceptible to multiple error types; whereas in classical error correction only bit-flips ($X$-errors) need to be considered, in quantum error correction an additional error type, phase-flips ($Z$-errors), must also be considered \cite{Devitt2013}.

The stabilizer framework for quantum error correction has been developed to allow quantum codes to be constructed within the above constraints \cite{Gottesman97,Gottesman:1998hu}. The essential idea is that the information encoded in a register of qubits is distributed across a larger entangled system of qubits. The extra degrees of freedom due to this expansion allows errors to do be detected using a series of `stabilizer measurements'. These stabilizer measurements reveal information about the parity of the register, whilst leaving the encoded quantum information unchanged. In general, a stabilizer code is labelled using the $[[n,k,d]]$ notation, where $n$ is the total number of qubits and $k$ is the number of logical qubits. The code distance $d$ is the minimum weight error that will go undetected by the code. We note also that the stabilizer framework and other error correction schemes can also be applied in a continuous time setting, where thermal dissipation can be used perform the corrections automatically (see e.g.~\cite{Pudenz14a,Jordan06a,sarovar05a,vinci15a,young13a,Sarovar13a}). However, this is not the subject of our current work.

The recently introduced coherent parity check (CPC) framework for quantum error correction \cite{chancellor16,roffe17,Roffe18a} provides a tool kit for the conversion of classical codes to quantum stabilizer codes. CPC codes are known to include large classes of stabilizer codes such as CSS codes, and are conjectured to be a superset of all known stabilizer codes. CPC codes have a specific structure in which the qubits are separated into two distinct types: data qubits and parity qubits. Error correction then proceeds via a three-part parity checking sequence. First, a set of bit parity checks are performed on the data qubits and the results copied to the parity qubits. This is followed by a second round of parity checks to detect phase-flips, with the results again being copied to the parity qubits. The final round of parity checks, referred to as cross-checks, takes place between the parity qubits themselves and fixes the code distance to the desired length. The specific advantage of the the CPC framework is that the parity checking sequences in each stage can be taken directly from existing classical codes.

The effect of a CPC encoder is to place the data and parity-check qubits in known stabilizer states, which can then be measured. Error propagation through such stabilizer measurements is identical to error propagation through a decoder (that is, the reverse of the encoder operations). One can therefore think either of stabilizer measurements on the encoded state, or, equivalently, the measurement of parity check qubits after a decoder. The results of all the stabilizer measurements can be combined to form a binary string called a syndrome. The role of a decoder in an error correction protocol is to infer the best recovery operation given the information provided by this syndrome. For classical codes, various methods exist that allow decoding to be performed efficiently  \cite{MacKay1996,MacKay1999}. Unfortunately, owing to the fact that stabilizer codes must be able to detect two-error types simultaneously, it is not always possible to use these existing techniques directly. As a result, bespoke decoding strategies need to be developed for use with quantum codes.

\subsection{The decoding problem}

Whether in a classical or quantum setting, the problem of decoding amounts to answering the same fundamental question: ``given the data available, what actions would be most likely to preserve the encoded information?'' There are generally recognized to be two different techniques \cite{Jaynes57a,Jaynes68a,Frieden72a} by which such inference can be achieved:

\begin{enumerate}
\item  Maximum likelihood estimation (MLE), for which it is assumed that the single most likely event, in this case the single most likely pattern of errors, has occurred.
\item Maximum entropy inference (MaxEnt), for which a distribution which maximizes entropy subject to constraints is used to draw probabilistic conclusions about the errors which make up the pattern.
\end{enumerate}

If all errors occur at the same rate and are less than $50\%$ likely to happen, MLE reduces to finding the the fewest number of errors which are consistent with the result of an error correction measurement (syndrome measurement).  On the other hand, MaxEnt relies on finding the probability of every event, so in principle will always depend in detail on the error rate(s). We discuss later how both of these approaches can be mapped to specialized hardware. 

Before continuing, it is worth considering a highly simplified example to demonstrate the difference between MLE and MaxEnt. Consider a simple (fictional) code where the observed parity measurements correspond to four distinct possibilities. In one of the four error patterns error $e_1$ is the only error, but in the other three patterns two errors other than $e_1$ have occurred. Based on MLE, we would always correct error $e_1$, since the single most likely error pattern (assuming all error rates are less than 50\%) is the one with the lowest weight. On the other hand the MaxEnt decoding depends on the error rate. Assuming all errors are equally likely and occur with a probability $p$, we can see that if $p<\frac{1}{3}$, then correcting $e_1$ will indeed reduce the error rate. However, if $\frac{1}{3}<e_1<\frac{1}{2}$, than it is more likely that the actual error pattern is one of the of the three cases in which error $e_1$ has \emph{not} occurred. In this case the MLE strategy actually does more harm than good, while the MaxEnt strategy does not perform the detrimental `correction'.

A practical decoding method is a key element in classical error correction code design. Leading classical error correction schemes, such as low density parity check (LDPC) codes and turbo codes, rely upon efficient approximate decoding strategies. For example, LDPC codes are decoded using belief propagation on a type of graphical model known as a factor graph \cite{MacKay1996,MacKay1999}.

A method for mapping quantum codes to a factor graphs has been demonstrated in \cite{Roffe18a}, but this technique required $Y$ errors to be represented as burst errors. In this work, as a by-product of our Ising model construction, we also prersent a factor graph construction which does does not require a multi-bit error model. While we focus on Ising model specialized hardware in this work, it is also worth noting that specialized hardware for belief propagation, for instance application specific circuits (ASICs), could also provide a promising route to improve decoding.

There is room for improvement in quantum decoding. As an example, a threshold of $p_c\approx10.3\%$ can be achieved on a 2D toric code using a minimum weight matching (a MLE technique) decoder \cite{Wang03a,Breuckmann17a}. However, statistical mechanics arguments based on an Ising model corresponding to these codes suggests a maximum threshold of $p_c\approx10.9\%$. Because this maximum threshold is calculated based on the statistical mechanical properties of an Ising model, a perfect thermal sampler on this model would be able to saturate the decoding bound \emph{by construction}. It is of course an open question whether such a high quality sampler could actually be practically implemented.

\subsection{Ising model computing}

Although it began as a model for magnetism \cite{Ising25a}, the Ising model has since been demonstrated to be a powerful tool in representing hard optimization and machine learning problems in specialized  hardware. Finding the ground state (lowest energy) of an Ising spin glass is known to be NP-hard. Therefore all NP-hard problems can be mapped to it with only polynomial overhead. Moreover, practical mappings of many important problems are actually known, with examples include partitioning, covering, and satisfiability \cite{Lucas2014a,Choi10a,Chancellor16c}. In addition to the ground state problem being NP-hard, Ising models are known to be universal in  the sense that \emph{any} classical spin model can be efficiently simulated by an Ising model \cite{De_las_Cuevas16a}. The algebraic similarity between Ising models and the decoding of codes was first noticed by Sourlas \cite{Sourlas1989a}, who demonstrated that under certain circumstances Ising spin glasses behave as optimal error correcting codes.

Probably the most well known specilized hardware which use Ising model based encodings are quantum annealers, such as those produced by D-Wave Systems Inc \cite{D-Wave}. Annealers, however, are not the only Ising model based computational machines. There have recently been efforts to produce other specialized Ising model based computing hardware \cite{McMahon16a,Inagaki16a,Yamaoka16a,Fujitsu_announce}, including some fully classical devices.

It is not only the ground states of Ising models which are computationally interesting. Thermal distributions over Ising models perform a constrained entropy maximization, which is useful for MaxEnt inference among other tasks. There has been much recent work for instance on how specialized annealing hardware may be used to realise Boltzmann machines by sampling a Boltzmann distribution \cite{Adachi2015,Amin2018a,Benedetti2016a,Benedetti2017a}. Additionally, Ising thermal distributions can be used to perform maximum entropy decoding of (classical) communications \cite{chancellor16b} via MaxEnt techniques.

Specialized Ising hardware can take an variety of forms, including fully classical CMOS devices which operate at room temperature \cite{Yamaoka16a,Fujitsu_announce}, and optimize over Ising models directly, rather than being arranged in a more traditional architecture. Using these devices directly for solid state quantum computing would mean a high rate of communication with a room temperature environment, and therefore a large heat and noise flux incident on the device. To avoid this problem, it would be preferable to perform the logical operations associated with decoding at deep crogenic, rather than room temperature.  CMOS information processing devices can be operated at deep cryogenic temperatures, and the importance for quantum computing has been highlighted \cite{Homulle2018a,Patra2018a,Weinreb2007a,Homulle2018b,Conway-Lamb2016a}. For estimates of power and area requirements for controls in silicon qubits, see \cite{Geck19a}. Given that quite complex devices, including field programmable gate arrays \cite{Homulle2018b,Conway-Lamb2016a} can be operated at cryogenic temperatures of around $4$ K, a cryogenic CMOS implementation of the (fully classical) Ising model computers being explored by \cite{Yamaoka16a,Fujitsu_announce} could provide a promising path for practically implementable quantum error correction.

\section{Mapping Quantum Error Correction to Ising Machines}

It has recently been demonstrated in \cite{Roffe18a} that the decoding of many quantum error correction codes can be mapped to a classical factor graph via the so called coherent parity check (CPC) formalism, originally introduced in \cite{chancellor16}. It is in turn known that due to the nature of the parity checks in these codes, their factor graphs naturally map to Ising models, which is the preferred encoding style of an emerging family of specialized computing hardware \cite{McMahon16a,Inagaki16a,Yamaoka16a,Fujitsu_announce}, including quantum annealers \cite{D-Wave,johnson11a}. This mapping is valid for all codes which can be described within the CPC framework. Whilst it is an open question as to which codes can be described within this framework, it is known that the framework can at least describe all CSS codes \cite{chancellor16}. It is also suspected that the CPC framework may be able to describe all stabilizer codes up to local unitaries. 

While we will not review the entire CPC formalism here, it is worth remarking on the key elements of the graphical construction, as it is important to understand how to map the decoding of quantum codes to Ising models. The construction of a classical factor graph using the graphical version of the CPC framework presented in \cite{Roffe18a} begins with a so called operational representation of the code, which represents how (unmeasured) data qubits interact with which (measured) parity check qubits. This operational representation must then be annotated with directed edges to represent indirect propagation of errors. As it is not directly relevant to the discussion here, we will not review how this annotation is done, but instead refer the reader to \cite{Roffe18a} for a simple set of graphical rules. 

In the operational representation, unmeasured data qubits are represented by triangles  \scalebox{.5}{\input{tikzit/data_qubit.tikz}} and measured parity check qubits are represented by stars \scalebox{.5}{\input{tikzit/parity_qubit.tikz}}. There are three types of edges between these qubits, representing the three ways errors can be transmitted: directly through bit error checks, directly through phase error checks, and through indirect propagation due to the quantum nature of the interactions. For the purposes of this paper, it is only necessary to review the translation of the qubits to classical factor graph representation and not how the edges translate. The data qubit translates to two classical data bits:
\begin{equation}
\input{tikzit/unmeasured_definition.tikz}.
\end{equation}
The parity check qubits on the other hand translates to a factor \footnote{We use the compressed factor graph notation used in \cite{Roffe18a}, where the single body elements of each factor are absorbed into the factor creating a `soft' constraint. This is consistent with the most compact and natural Ising model representation. While this differs from the traditional Tanner graph representation, it is unambiguous and a translation between the two consists of converting the factor to a hard constraint and adding an additional bit with the weight of the factor. }, representing the measured bit information, and a classical data bit representing the unmeasured phase information:
\begin{equation}
\input{tikzit/measurement_definition.tikz}.
\end{equation}
Finally, it is possible for a parity check bit to have a self loop, where its own phase information is propagated to its (measured) bit degree of freedom. In this case, the qubit again translates to a parity check node and a classical bit, but this time with an edge connecting them,
\begin{equation}
\input{tikzit/measurement_definition_self.tikz}.
\end{equation}
The classical factor graphs which represent these quantum error correcting codes will have two types of nodes: classical bits representing whether or not a degree of freedom was errored, and parity check nodes representing measurements of the bit degrees of freedom of the parity check qubits. These measurements effectively tell us the parity of errors (whether an odd or even number have occurred) over a given set of bit and phase degrees of freedom. If we assume a very simple error model where all qubits have an equal likelihood to have a bit (X) or phase (Z) error, and no other error types are allowed, then the energy with respect to the following Ising Hamiltonian is proportional to the number of errors up to an irrelevant offset
\begin{equation}
H^{\mathrm{err\,count}}=-\sum_{i=1}^{k+n} \sigma^z_i-\sum_{j=1}^{n-k}s_j\prod_{l \in Q_j}\sigma^z_l
\end{equation}
where $s_j\in \{+1,-1\}$ is the parity of the $j$th error measurement and $Q_j$ is the set of degrees of freedom (represented as classical bits) checked by that measurement. In this case, an Ising spin taking a $+1$ value indicates no error, while a $-1$ value indicates that an error has occurred. For the bit degrees of freedom of the parity checking qubits, a value of the parity checking measurement which is different from $s_j$ indicates an error. Since an $X$ or $Z$ error on any qubit increases the energy with respect to this Hamiltonian by $2$, it follows that this Hamiltonian effectively counts the number of errors necessary to give a syndrome measurement $\{s\}$, and further follows that the lowest energy state is the one with the fewest errors. Furthermore, if we assume that our errors have a probability less than $50\%$, than the error configuration(s) with the least errors are also the most likely. Finding the ground state of this Hamiltonian is therefore equivalent to performing maximum likelihood (MLE) inference.

This Hamiltonian, however, is not only a tool for MLE inference. Let us consider a Boltzmann distribution
\begin{equation}
p_a=\frac{\exp\left(-\frac{E_a}{T}\right)}{Z} \label{BD}
\end{equation}
where $E_a$ is the energy of configuration $a$, i.e. the value of the Hamiltonian, and $Z$ is the partition function which guarantees that the probability distribution is normalized. Recall that $E_a$ is proportional to the number of errors required to measure the syndrome given by $\{s\}$ with the error configuration $a$ on the unmeasured degrees of freedom up to an energy offset. The structure of the Boltzmann distribution means that for every additional error the probability of the configuration is decreased by a multiplicative factor of $\exp(-\frac{2}{T})$. Since the probability of an error configuration in the actual decoding process is also decreased by a constant multiplicative factor if the number of errors in the configuration is increased by one, it follows that sampling from the Boltzmann distribution over the Ising model is the same as sampling over error distributions at a finite error rate. Inferring the likelihood of specific errors at a known overall error rate is a form of maximum entropy (MaxEnt) inference, a powerful tool which allows knowledge of the error rate to help with decoding. There is a formally rigorous relationship between the error rate and the temperature of the distribution, namely, sampling at an error rate $p$ is equivalent to sampling the Boltzmann distribution at the Nishimori temperature \cite{Nishimori1980,Nishimori2001}
\begin{equation}
T_{Nish}=2\,\left(\ln\left(\frac{1-p}{p}\right)\right)^{-1}.
\end{equation}
If the error rate is known, at least approximately, as it almost always will be on real devices, then maximum entropy inference can use this additional information to perform better than maximum likelihood. One dramatic example of this is \cite{chancellor16b} where maximum entropy inference performed on a programmable quantum annealer was shown to be able to out-perform perfect maximum likelihood decoding. 

While the example given here provides motivation for the possibility of using Ising machine based techniques, including maximum entropy inference to decode quantum error correcting codes, it has thus far been done in a rather unrealistic setting. Assuming equal error rates on every degree of freedom is natural in classical communications, where parity checks and data will all be sent through the same channel. The same assumption is somewhat artificial in the setting of quantum error correction, where there is no a priori reason to expect that bit and phase errors will occur at the same rate. Furthermore, we have thus far considered an error model which does not contain a separate mechanism for $Y$ errors, where the bit and phase degrees of freedom on a qubit flip preferentially at the same time. While these errors can be treated as burst errors in the factor graph, there is no obvious natural way to include burst errors directly in the Ising Hamiltonian. We show in the next section how arbitrary single qubit error models can be represented as Ising models, and in turn how they can be represented as weighted factor graphs where burst errors need not be explicitly included to represent arbitrary single qubit error models.

\subsection{Deducing the error pattern from Ising decoding}

We now briefly discuss how the logical error pattern can be deduced from the Ising model. This can be done in two ways. Firstly for MaxEnt the error pattern is deduced by examining the lowest energy configuration found (or one of the lowest energy configurations chosen at random in the event of a tie) and calculating what logical corrections are needed. For MaxEnt, given a collection of configurations sampled with Boltzmann weighted probabilities, logical errors can be deduced by taking a `vote' of all sampled configurations for each qubit, i.e. by tabulating the number for which no correction, an $X$ correction, a $Z$ correction, or a $Y$ correction is appropriate. The correction with the most `votes' is the one which is most likely to correct the information when averaging over all possible error patterns. If information about other kinds of errors is required, then a similar approach can be followed.

\subsection{Hamiltonian terms for $Y$ errors}

Recall that the maximum likelihood error configuration(s) needed for MLE decoding are recovered from MaxEnt (maximum entropy) inference by isolating the most probable (lowest energy) configuration(s). This can be shown directly by observing that at a sufficiently low temperature, the Boltzmann distribution on an Ising model will be dominated by the lowest energy configuration(s). It is therefore sufficient to demonstrate an Ising model construction where the Boltzmann distributions can be used to perform MaxEnt inference for arbitrary $X$, $Y$, and $Z$ error rates, as such a model will necessarily also be able to perform MLE inference by focusing only on the lowest energy states (equivalent to a $T=0$ Boltzmann distribution). 

We need to consider both errors on the data qubits and the parity check qubits. Because it is conceptually simpler, we start with the case of a data qubit. In the factor graph representation, such a qubit is represented by two classical bit variables, one for the bit information of the qubit and one for the phase information. These bits can be represented as Ising spins. In an Ising model representation, these two spins can have a coupling between them. We therefore write the total Hamiltonian as follows
\begin{equation}
H^{\mathrm{data}}=-h_1\sigma^z_1-h_2\sigma^z_2-J\,\sigma^z_1\sigma^z_2. \label{H_data}
\end{equation}

To map a full $X$, $Z$, $Y$ error model we must map the probabilities of these spins being in the $+1,+1$, $-1,+1$, $+1,-1$ or $-1,-1$ state to the no error, bit flip error only ($p_x(1-p_z)$), phase flip error only ($p_z(1-p_x)$), and bit and phase flip error ($p_{xz}=p_xp_z+p_y$) probabilities respectively for a Boltzmann distribution \eqref{BD} obtained for this Hamiltonian at a finite temperature $T$.

To avoid having to calculate the partition function, we compare ratios of probabilities between the Boltzmann distribution and the distribution we are trying to emulate, for instance
\begin{eqnarray}
\frac{p_{(-1,+1)}}{p_{(+1,+1)}}&=&\exp\left(-2\frac{h_1+J}{T}\right) \nonumber \\
&=&\frac{p_x(1-p_z)}{1-p_x-p_{xz}-p_z+2p_xp_z}\equiv\bar{p}_x \label{data_x}
\end{eqnarray}
where the definition of $\bar{p}_x$ is for mathematical convenience in later calculations. We furthermore set
\begin{eqnarray}
\frac{p_{(+1,-1)}}{p_{(+1,+1)}}&=&\exp\left(-2\frac{h_2+J}{T}\right)\nonumber \\
&=&\frac{p_z(1-p_x)}{1-p_x-p_{xz}-p_z+2p_xp_z}\equiv\bar{p}_z, \label{data_z}
\end{eqnarray}
and
\begin{eqnarray}
\frac{p_{(-1,-1)}}{p_{(+1,+1)}}&=&\exp\left(-2\frac{h_1+h_2}{T}\right)\nonumber \\
&=&\frac{p_{xz}}{1-p_x-p_{xz}-p_z+2p_xp_z}\equiv\bar{p}_{xz}. \label{data_xz}
\end{eqnarray}
We now have three equations and three unknowns, solving for the terms in Eq. \ref{H_data}, we find that,
\begin{eqnarray}
 J&=&\frac{T}{4}\left[-\ln(\bar{p_x})-\ln(\bar{p_z})+\ln(\bar{p}_{xz})\right] \nonumber \\
 h_1&=&-\frac{T}{2}\,\ln(\bar{p}_x)-J \nonumber \\
 h_2&=&-\frac{T}{2}\,\ln(\bar{p}_z)-J.
\end{eqnarray}
Now we turn to the parity checking qubits, each of which is represented by only a single bit corresponding to the phase degree of freedom. Whether or not the bit degree of freedom has been errored can be inferred by checking whether the measured parity value agrees with the parity of the errors it detected. Stated another way, if we deduce that an odd (even) number of bits in the parity check were errored and  the parity check shows and even (odd) parity, then we conclude that the bit degree of freedom was corrupted. As discussed in the previous section, we encode the measured error value in the sign of the parity check over the bits $Q$. If no error has been detected, then the coupling strength in the Ising mapping will be negative (ferromagnetic). Otherwise it will be positive (anti-ferromagnetic). To take into account the effects of $Y$ errors on the parity check as well as the phase degree of freedom, we introduce an additional multi-body term which acts on the phase bit ($b_p$) as well as the bits within the parity check \footnote{Or, in the case where the phase bit is already contained in $Q$, we add a new term which acts on all of the bits in the parity check \emph{except} for the phase bit}. The resultant Hamiltonian parity check term takes the form
\begin{eqnarray}
H^{\mathrm{par}}_\pm(Q,b_p)=-h_{\pm}\sigma^z_{b_p}-a_{Q\pm} \prod_{i\in Q}\sigma^z_i - \nonumber \\
- a_{Q\,b_p\pm} \sigma^z_{b_p} \prod_{i\in Q}\sigma^z_i .
\end{eqnarray}
As was done for the data qubits, focusing first on the $+$ case, where an error is not detected, we solve for ratios of probabilities
\begin{align}
&\frac{p_{(-1,+1)}}{p_{(+1,+1)}}=\exp(-2\frac{a_{Q+}+ a_{Q\,b_p+} }{T})=\bar{p}_x \nonumber \\
&\frac{p_{(+1,-1)}}{p_{(+1,+1)}}=\exp(-2\frac{h_++ a_{Q\,b_p+}}{T})=\bar{p}_z \nonumber \\
&\frac{p_{(-1,-1)}}{p_{(+1,+1)}}=\exp(-2\frac{h_++ a_{Q+}}{T})=\bar{p}_{xz}.
\end{align}
These equations have the same mathematical structure as Eqs. \eqref{data_x}, \eqref{data_z}, and \eqref{data_xz}, it follows immediately that 
\begin{eqnarray}
 a_{Q\,b_p+}&=&\frac{T}{4}\left[-\ln(\bar{p_x})-\ln(\bar{p_z})+\ln(\bar{p}_{xz})\right] \nonumber \\
 a_{Q+}&=&-\frac{T}{2}\,\ln(\bar{p}_x)-a_{Q\,b_p+} \nonumber \\
 h_+&=&-\frac{T}{2}\,\ln(\bar{p}_z)-a_{Q\,b_p+}.
\end{eqnarray}
The task now remains to find the values for the $-$ case, in other words when an error has been detected. In this case we have 
\begin{eqnarray}
\frac{p_{(-1,+1)}}{p_{(+1,+1)}}&=&\exp(-2\frac{a_{Q-}+a_{Q\,b_p-} }{T})=\bar{p}_x^{-1} \nonumber \\
\frac{p_{(+1,-1)}}{p_{(+1,+1)}}&=&\exp(-2\frac{h_-+ a_{Q\,b_p-}}{T})=\bar{p}_{xz}\bar{p}_x^{-1} \nonumber \\
\frac{p_{(-1,-1)}}{p_{(+1,+1)}}&=&\exp(-2\frac{h_-+a_{Q-}}{T})=\bar{p}_{z}\bar{p}_x^{-1}.
\end{eqnarray}
Except for the difference in the RHS, these equations again have the same mathematical structure. We can solve for the relevant terms as follows 
\begin{eqnarray}
 a_{Q\,b_p-}&=&\frac{T}{4}\left[\ln(\bar{p}_{x})-\ln(\bar{p}_{xz})+\ln(\bar{p}_{z}))\right] \nonumber \\
 a_{Q-}&=&\frac{T}{2}\,\ln(\bar{p}_x)-a_{Q\,b_p-} \nonumber \\
 h_-&=&-\frac{T}{2}\,\left[\ln(\bar{p}_{xz})-\ln(\bar{p}_{x})\right]-a_{Q\,b_p-}.
\end{eqnarray}
With the above, we have now completed all of the necessary terms to construct a complete Ising Hamiltonian for which the Boltzmann distribution performs maximum entropy inference on a general quantum error correcting code. In the next subsection we discuss the full Hamiltonian, as well as its expression as a factor graph.

\subsection{Full Hamiltonian and factor graph representation}

Now that we have constructed all of the necessary pieces, lets construct the whole Hamiltonian for maximum entropy inference on a CPC code. This Hamiltonian will involve $k$ Ising spins which represent the bit degrees of freedom on the data qubits $b$, $k$ spins which represent the phase information $p$, and $n-k$ spins which represent the phase degrees of freedom on parity check qubits. The Hamiltonian takes the form
\begin{eqnarray}
H_{\mathrm{decode}}&=&\sum^k_{i=1}(-h_1\sigma^z_{b_i}-h_2\sigma^z_{p_i}-J\sigma^z_{b_i}\sigma^z_{p_i}) + \nonumber \\
&&+\sum^{n-k}_{j=1}H^{\mathrm{par}}_{s_j}(Q_j,p'_j),
\end{eqnarray}
where $Q_j$ is the set of degrees of freedom (represented by Ising spins) checked by the $j$th parity checking qubit, and $s_j\in\{+1,-1\}$ is the syndrome measurement for that qubit. 

Because this Hamiltonian has been constructed so that Boltzmann distributions with respect to it reproduce error probabilities, the probability of an individual error configuration will be proportional to $\exp(-E/T)$, the exponent of that configuration's energy. We therefore deduce by the fact that this function decreases monotonically, that the lowest energy configuration with respect to this Hamiltonian is also the single most likely one. Therefore for a Hamiltonian constructed at \emph{any} finite positive temperature $T$, the maximum likelihood configuration is the ground state.

In addition to being represented as a Hamiltonian, the decoding problem including $Y$ errors can also be represented as a (weighted) factor graph. This can be done by examining the interactions in the Hamiltonian and constructing a graphical model from them. This model is similar to the original factor graph considered in \cite{Roffe18a}, but has one additional factor per qubit to adjust the error probabilities to include $Y$. The rules given in \cite{Roffe18a} can be modified as follow to explicitly include a $Y$ error in the factor graph, rather than treating it as a burst error on the code. The first modification is that when mapping a data qubit (represented by a triangle, \scalebox{.5}{\input{tikzit/data_qubit.tikz}} in the graphical language of that paper), an additional weighted factor needs to be added to include correlations between the bit and phase degrees of freedom of that qubit. The translation of a data qubit in the operational representation is therefore
\begin{equation}
\input{tikzit/unmeasured_definition_Y.tikz},
\end{equation}
where the color of the factor is added to emphasize that it is weighted differently \footnote{or if considered as a hard constriant that the isolated data bit which the node connects to is weighted differently.} from the parity checks. For the parity check qubits, (denoted by a star \scalebox{.5}{\input{tikzit/parity_qubit.tikz}}) two separate cases need to be considered, both the cases with and without a self loop. Without the self loop the definition is 
\begin{equation}
\input{tikzit/measurement_definition_Y.tikz},
\end{equation}
where edges extending off the edge of the figure correspond to the parity checks which the qubit performs on other qubits. The magenta color indicates the additional weighted factor to take the $Y$ errors into account. Edges representing other qubits checking the phase qubit have been omitted for visual clarity. The final case we need to consider is the case where the parity check qubit has a self-loop in the annotated operational representation. In this case the additional factor interacts with all of the qubits the parity qubit checks, but not the phase bit
\begin{equation}
\input{tikzit/measurement_definition_Yself.tikz}.
\end{equation}
Edges representing other qubits checking the parity check qubit have again been omitted for visual clarity. Aside from these variations of the node definitions, the explicit factor graph construction for a model including $Y$ errors is exactly the same as the one given in \cite{Roffe18a}.

\subsection{Decoding over time}

The foregoing discussion has implicitly assumed that the gates comprising the decoder/stabilizer measurements do not themselves introduce error. To be fully general, all quantum decoders must be able to tolerate gate errors. The standard solution in QEC to process syndrome data fully fault-tolerantly is for decoding to happen over time as well as space (see for example \cite{Dennis2001} in the context of surface codes). The idea here is that not only is the syndrome measurement at one time used to determine the likely event, but rather an entire sequence of syndrome measurements is used to reconstruct the most likely series of events. The Ising and factor graph approaches can be extended to this kind of decoding as well. If we first consider that each qubit starts out in the unerrored state, we begin with the standard factor graph or Ising model for the single-shot decoding from the first round of syndrome gathering. For the second round however, rather than single body terms corresponding to whether or not a qubit has been errored, another copy of the Ising model description of the code should be added. However, rather than having single body terms, it should have couplings to the previous copy. If an error has occurred in the previous round and persists, it should trigger the same syndromes. The single body terms on data qubits thus become two body terms coupling to the previous round, and likewise the two body terms become four body terms. The terms relating to parity check qubits remain unchanged, as syndrome information is measured and not carried forward from round to round. 

Mathematically, the transformation from single round decoding to decoding over time transforms Eq.~\ref{H_data} to a comparison with the values observed at the previous time,
\begin{eqnarray}
H^{\mathrm{data}}=h_1\sigma^z_{1,t-1}\sigma^z_{1,t}+h_2\sigma^z_{2,t-1}\sigma^z_{2,t}- \nonumber \\
-J\,\sigma^z_{1,t-1}\sigma^z_{2,t-1}\sigma^z_{1,t}\sigma^z_{2,t}, \label{H_data_time}
\end{eqnarray}
where $t$ and $t-1$ are indices to indicate slices representing syndrome measurements at different times. The quantities $h_1$, $h_2$, and $J$ are the same as in Eq.~\ref{H_data} because the statistics of the errors is agnostic to how the correction is performed. As with decoding at a single time, the interactions on a single data qubit can be expressed as a factor graph, with a pair of bits per measurement cycle, 
\begin{equation}
\input{tikzit/over_time.tikz}.
\end{equation}
Each factor is labelled with its weight, and edges which do not connect to vertexes indicate interactions with the parity checks within a time slice. Note that except for the leftmost set of classical bits, there are no single body terms.

\section{Numerical results}

\begin{figure}
\includegraphics[width=6cm]{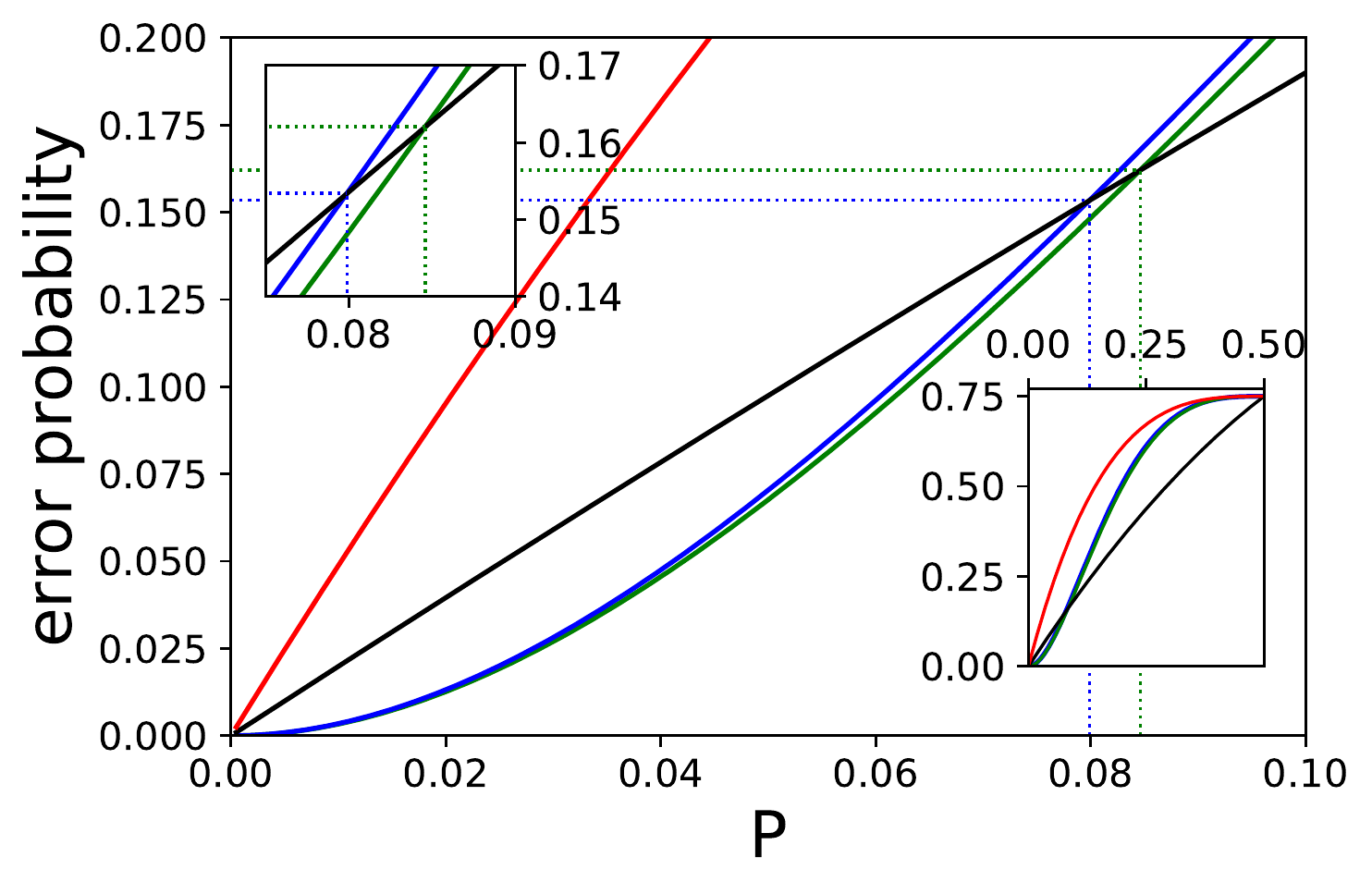}
\begin{centering}
\caption{\label{fig:5_1_3_thresh} Probability of a logical $X$, $Z$ or $Y$ in a $[[5,1,3]]$ code for $p_x=p_z=p$ and $p_y=0$. Blue is maximum likelihood decoding, green is maximum entropy, and red is for data which is encoded but no corrective action is taken. The black line is the raw averaged error probability of an unprotected qubit subject to the same errors. Dotted lines are guides to the eye to show the threshold under the two decoding strategies. The upper inset is a zoom on the threshold while the lower inset shows a wider range of $p$.}
\par
\end{centering}
\end{figure}

We now compare numerical results of maximum entropy and maximum likelihood inference for two codes, a $[[5,1,3]]$ code and a $[[9,3,3]]$ code, the CPC matrices for which can be found in the appendix. We choose small codes where the probabilities can be calculated exhaustively and so that there is no statistical error in our calculations and thresholds can readily be calculated by bisection. We consider a simple error model with a single round of error correction and perfect syndrome collection. This is appropriate as the goal of these calculations is to provide proof-of-principle for our decoding methods, not to demonstrate the practicality of the codes.

 For the  $[[5,1,3]]$ code, we consider a situation where $p_x=p_z=p$ and $p_y=0$, in other words equal probability of dephasing or bit flip errors, but no error which independently implements a combined flip and dephasing error. The results of this decoding can be found in Fig.~\ref{fig:5_1_3_thresh}. This figure demonstrates that not only does the  maximum entropy strategy reduce the probability of a logical error compared to maximum likelihood, it also  moves the threshold, defined as the point for which the probability of an error exceeds the probability for an unprotected qubit, by approximately $0.5\%$. For a summary of threshold values, see table  \ref{tab:thresholds}.

\begin{figure}
\includegraphics[width=6cm]{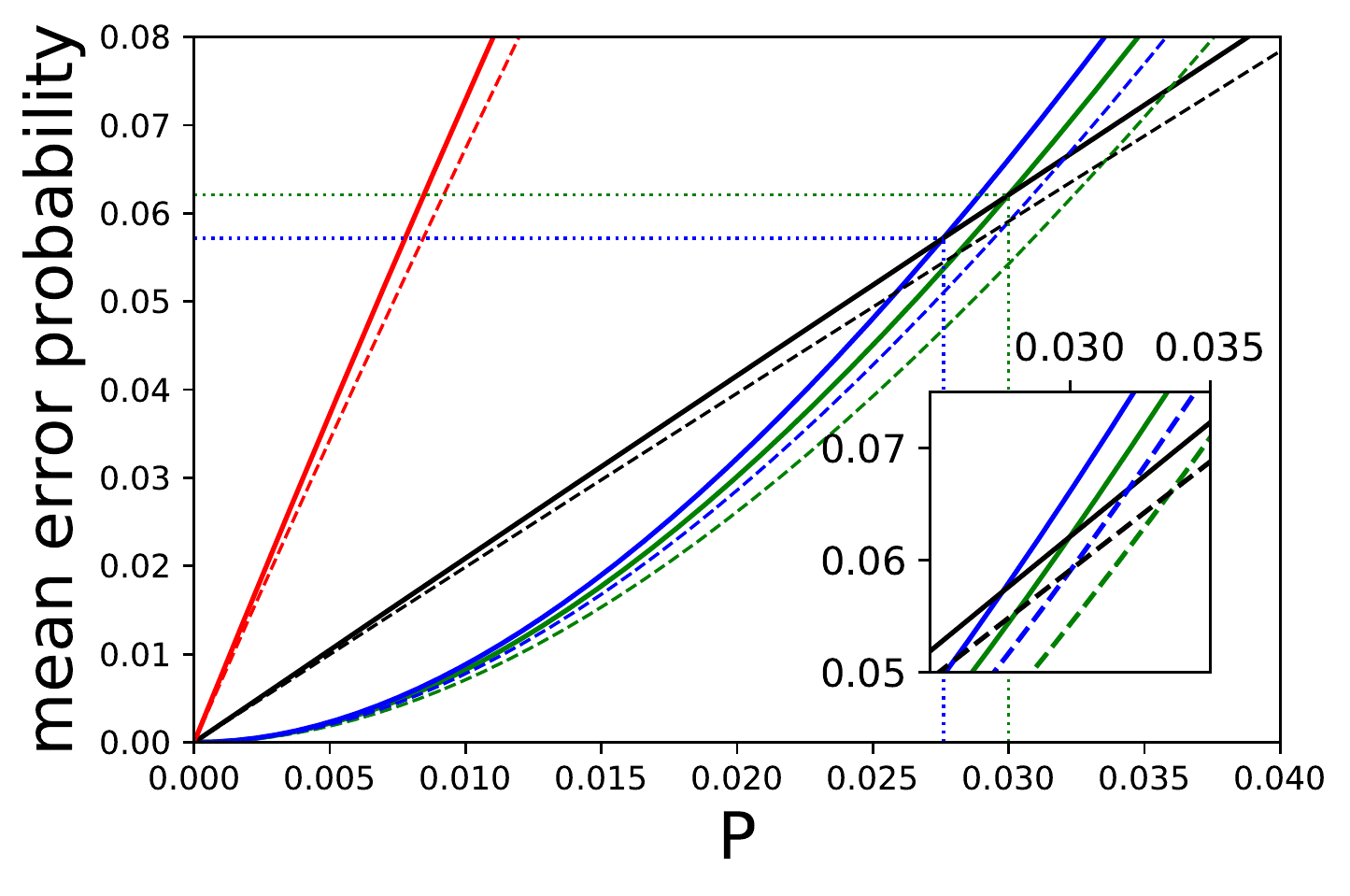}
\begin{centering}
\caption{\label{fig:9_3_3_thresh} Probability of a logical $X$, $Z$ or $Y$ averaged over the logical degrees of freedom for a $[[9,3,3]]$ code for $p_x=p_z=p$ and $p_y=0.1 p$ (rates with no independent $Y$ error are depicted as dashed lines for comparison). Blue is maximum likelihood decoding, green is maximum entropy, and red is for data which is encoded but no corrective action is taken. The black line is the raw averaged error probability of an unprotected qubit subject to the same errors. Dotted lines are guides to the eye to show the threshold under the two decoding strategies. The inset is a zoom on the threshold.}
\par
\end{centering}
\end{figure}

As a more sophisticated example, we consider a $[[9,3,3]]$ code with $p_x=p_z=p$ and $p_y=0.1 p$, the results for which are shown in Fig.~ \ref{fig:9_3_3_thresh}. This more sophisticated multi-qubit code again shows better decoding from a maximum entropy strategy and a shift in the threshold. All threshold values are depicted in table \ref{tab:thresholds}. 

\begin{figure}
\includegraphics[width=6cm]{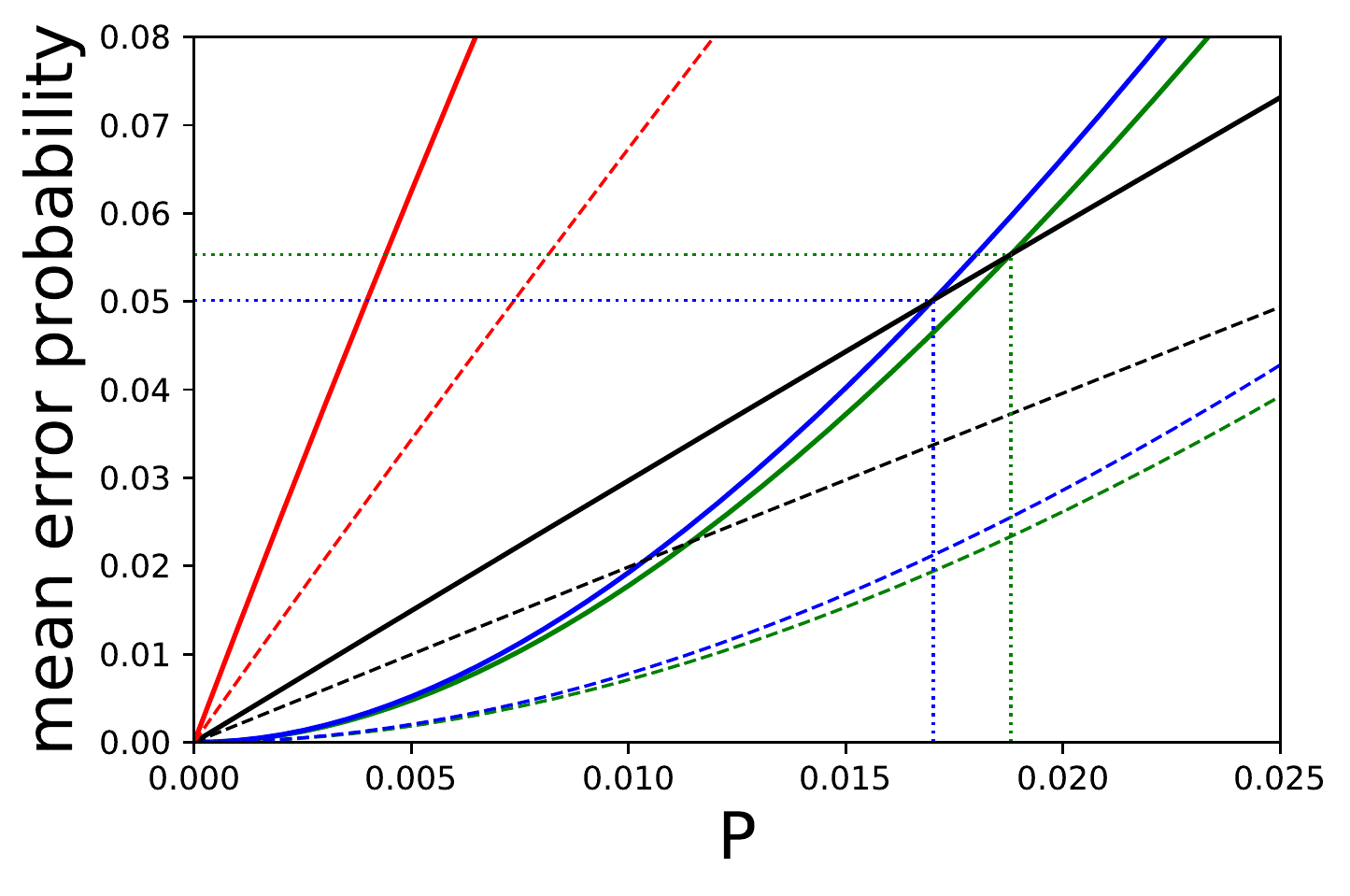}
\begin{centering}
\caption{\label{fig:9_3_3_iso} Probability of a logical $X$, $Z$ or $Y$ averaged over the logical degrees of freedom for a $[[9,3,3]]$ code for $p_x=p_z=p$ and $p_y=p-2\,p^2$, corresponding to isotropic errors (rates with no independent $Y$ error are depicted as dashed lines for comparison). Blue is maximum likelihood decoding, green is maximum entropy, and red is for data which is encoded but no corrective action is taken. The black line is the raw averaged error probability of an unprotected qubit subject to the same errors. Dotted lines are guides to the eye to show the threshold under the two decoding strategies.}
\par
\end{centering}
\end{figure}

One final case which is often considered theoretically is the case of completely isotropic error, $p_x=p_z=p_{xz}=p$. Solving for $p_y$ we find that this case corresponds to $p_y=p-2\,p^2$. The results for the $[[9,3,3]]$ code with isotropic error appear in Fig.~\ref{fig:9_3_3_iso}.

\begin{table*}
\begin{tabular}{|c|c|c|c|}
\hline 
code type & MLE threshold & MaxEnt threshold & difference\tabularnewline
\hline 
\hline 
$[[5,1,3]]$, $p_{y}=0$ & $p=0.07989$ & $p=0.08460$ & $0.004708$\tabularnewline
\hline 
$[[9,3,3]]$, $p_{y}=0$ & $p=0.03005$ & $p=0.03358$ & $0.003534$\tabularnewline
\hline 
$[[9,3,3]]$, $p_{y}=0.1p$ & $p=0.02760$ & $p=0.03000$ & $0.002401$\tabularnewline
\hline 
$[[9,3,3]]$, $p_{y}=p-2\,p^2$ & $p=0.01701$ & $p=0.01880$ & $0.001789$\tabularnewline
\hline 
\end{tabular}

\caption{\label{tab:thresholds} Thresholds of maximum entropy (MaxEnt) and maximum likelihood (MLE) strategies for different codes considered in this paper.}

\end{table*}

\section{Implementation}

The Ising model representation of quantum error correcting codes has a fixed structure, in other words, the measured syndrome values $\{s\}$ affect the magnitude and sign of different coupling elements, but not which variables interact with which other variables. This structure means that any decoder could be constructed in an \emph{application specific} way, much in the way that an application specific integrated circuit (ASIC) approach is used in some cases in electrical engineering. In fact for superconducting circuit annealers and many quantum inspired implementations, the decoder itself would actually be an ASIC, but in the interest of generality, we refer to hardware which is designed to only perform quantum error correction as an application specific processor. 

The application specific approach affords several advantages over using general purpose Ising hardware. Probably most importantly, the hardware can be designed to minimize or eliminate the costs associated with embedding the problem. It is generally recognized that the embedding overheads are a major obstacle for current quantum annealer technology \cite{Biswas17a,Rieffel15a,Hammerly18a}. For a quantum annealer, this would require multi-body interactions, however there are several ideas currently in the literature on how this can be done in hardware \cite{Chancellor17b,Leib16a,Rocchetto201a,Lechner2015a,Zhao17a,MelansonAQC2018}.  It could also be done using a variety of mapping techniques from multi-body to two-body couplers (a process known as `quadratization'). For a review of these techniques see \cite{Dattani18a}. Additionally, if the decoding is made with the same type of hardware as the quantum computer it is correcting (superconducting circuits or trapped atoms/ions for instance), it is likely that the connectivity constraints on the two devices will be similar and that an efficient embedding of the decoding problem can be found.

An additional benefit of an application specific approach is that the control requirements would be much less demanding, as the device would only have to be able to have a number of user controlled binary variables equal to the number of syndromes, rather than to approximate a continuum of strengths for each qubit and coupler. On currently implemented quantum annealing hardware, the area taken up by the classical control digital to analog converters (DACs) can be a limiting factor in chip design \cite{Bunyk2014a}, reducing the the complexity of the controls would free up area for qubits and couplers and allow for more powerful devices.

Another benefit of using specialized hardware to decode is that the decoding hardware, whether quantum or classical, can be directly integrated with the quantum computer it is correcting. For a superconducting circuit architecture this would mean having quantum and/or classical error correction hardware in the same cryostat, thus cutting down on requirements to communicate with devices at room temperature and reducing the heat flux incident on the device. This approach could also reduce communication latency, and potential delays waiting for results to be returned from more traditional classical computing software. 

\begin{figure}[t]
\begin{centering}
\includegraphics[width=8cm]{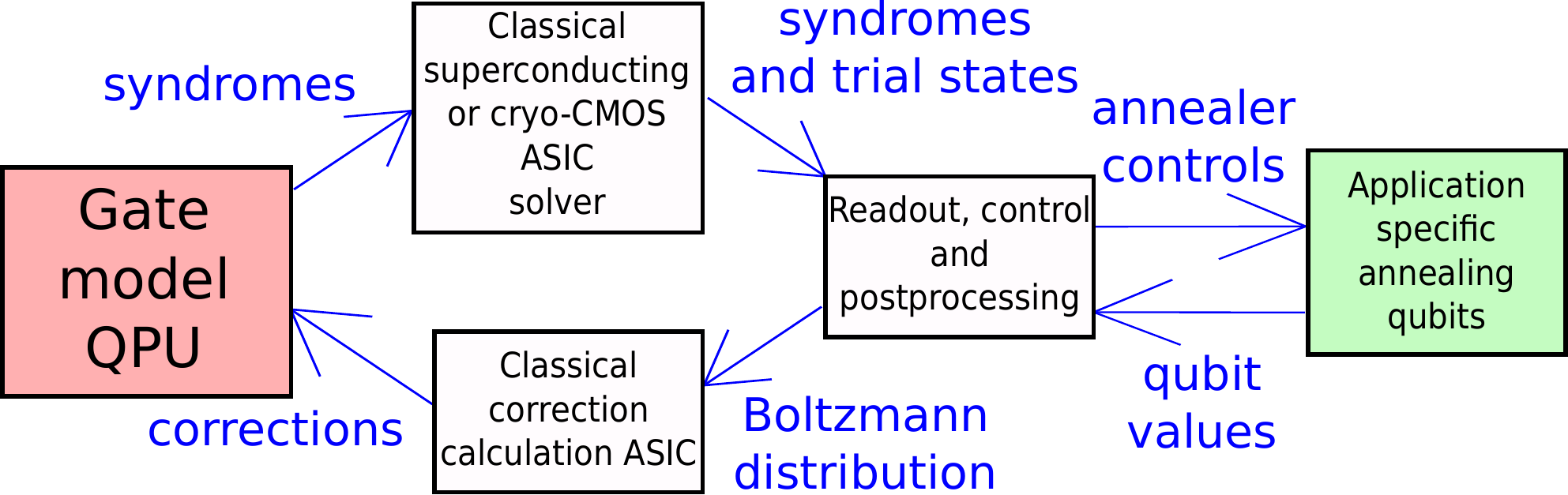}
\end{centering}
\caption{ Information flow in a hybrid quantum/classical annealer accelerated error correction setup. Classical hardware finds high probability error configuration which are then converted an approximate sampling of a Boltzmann distribution with the aid of a quantum annealer.  \label{fig:ASIC_annealer_block}}
\end{figure}

Decoding could also potentially be performed using hybrid quantum/classical techniques, for instance a superconducting quantum annealer could be paired with a cryo-CMOS or classical superconducting ASIC which uses belief propagation, or other classical techniques to find high quality solutions, but which are not distributed in a Boltzmann distribution. Reverse annealing \cite{Chancellor17a,RA_whitepaper,Perdomo-Ortiz11} could then use these as a starting point to find an approximate Boltzmann distribution for use in MaxEnt inference. In particular, it has recently been shown that two orders of magninitude improvement in solution time is possible through reverse annealing even when starting from a solution found by a simple greedy search \cite{Venturelli18a}. A graphical representation of information flow in an application specific annealer accelerated error correction unit appears in Fig.~\ref{fig:ASIC_annealer_block}.

\section{Discussion and outlook}

In this work we have developed a general way to map the problem of decoding quantum error correction to specialized Ising model hardware, which works for any code describable in the CPC formalism (at least all CSS codes, if not all stabilizer codes). Our methods are general in the sense that depending on the capabilities of the Ising hardware they can perform either maximum likelihood or maximum entropy decoding. We have demonstrated numerically that maximum entropy decoding using our formalism can improve the decoding of a small code. 

Since large scale quantum annealers actually exist, it would be feasible to test the performance of these devices with simulated large scale quantum error correction. Similar studies have already been performed for classical error correction in \cite{chancellor16b}, and an extension to more complex codes is forthcoming \cite{DWclassicalECinPrep}. The choice to map small codes in this paper was only made for presentation and to make the exact thermal distribution calculable. The complexity of our mapping from quantum codes to Ising models scales mildly with code size, and therefore would be feasible even for very large quantum error correction codes.

\acknowledgements 
JR acknowledges the support of the QCDA project which
has received funding from the QuantERA ERA-NET Cofund in Quantum Technologies implemented within the European Union's Horizon 2020 Programme. SZ acknowledges support from the Oxford-Man Institute for Quantitative Finance. DH acknowledges support from the ``Investissements d'avenir'' (ANR-15-IDEX-02) program of the French National Research Agency.  NC acknowledges support from from EPSRC grant refs EP/L022303/1 and EP/S00114X/1.

\bibliography{decode-bib.bib}

\section*{Appendix: codes used in numerical examples}

$[[5,1,3]]$ code

\begin{equation}
M_b=\left(\begin{array}{c}
0 \\
0 \\
1 \\
1
\end{array}\right)\,\,\,\,
M_p=\left(\begin{array}{c}
1 \\
0 \\
0 \\
1
\end{array}\right)\,\,\,\,
M_c=\left(\begin{array}{cccc}
0 & 1 & 0 & 1 \\
0 & 0 & 1 & 1 \\
0 & 0 & 0 & 1 \\
0 & 0 & 0 & 0
\end{array}\right)
\end{equation}

$[[9,3,3]]$ code

\begin{equation}
M_b=\left(\begin{array}{ccc}
0 & 1 & 1 \\
1 & 0 & 0 \\
1 & 0 & 1 \\
1 & 0 & 1 \\
0 & 1 & 1 \\
0 & 0 & 1
\end{array}\right)\,\,\,\,
M_p=\left(\begin{array}{ccc}
1 & 1 & 1 \\
1 & 0 & 1 \\
0 & 0 & 0 \\
1 & 1 & 0 \\
0 & 0 & 0 \\
1 & 1 & 1
\end{array}\right)
\end{equation}

\begin{equation*}
M_c=\left(\begin{array}{cccccc}
0 & 1 & 0 & 0 & 1 & 0 \\
0 & 0 & 1 & 0 & 0 & 0 \\
0 & 0 & 0 & 0 & 1 & 0 \\
0 & 0 & 0 & 0 & 0 & 1 \\
0 & 0 & 0 & 0 & 0 & 0 \\
0 & 0 & 0 & 0 & 0 & 0
\end{array}\right)
\end{equation*}

\end{document}

%% file: tikzit/tikz_premable.tex
\usepackage[svgnames]{xcolor}
\usepackage{tikz}
\usetikzlibrary{decorations.markings}
\usetikzlibrary{shapes.geometric}

\pgfdeclarelayer{edgelayer}
\pgfdeclarelayer{nodelayer}
\pgfsetlayers{edgelayer,nodelayer,main}

\tikzstyle{none}=[inner sep=0pt]

\tikzstyle{rn}=[circle,fill=Red,draw=Black,line width=0.8 pt]
\tikzstyle{gn}=[circle,fill=Lime,draw=Black,line width=0.8 pt]
\tikzstyle{yn}=[circle,fill=Yellow,draw=Black,line width=0.8 pt]
\tikzstyle{data_qubit}=[regular polygon,regular polygon sides=3,shape border rotate=0,fill=RoyalBlue,draw=Black]
\tikzstyle{parity_check_qubit}=[shape=star,star points=5, star point ratio=2.25,fill=Red,draw=Black]
\tikzstyle{hard_parity_constraint}=[rectangle,fill=White,draw=Black,scale=2,label={[xshift=0 cm, yshift=-0.5 cm]+}] 
\tikzstyle{classical_parity_check}=[rectangle,fill=White,draw=Black,scale=2] 
\tikzstyle{classical_parity_checkY}=[rectangle,fill=White,draw=Magenta,scale=2] 
\tikzstyle{classical_data_bit}=[circle,fill=Gold,draw=Black,scale=0.8]
\tikzstyle{classical_data_phase}=[circle,fill=DeepSkyBlue,draw=Black,scale=0.8]
\tikzstyle{classical_data}=[circle,fill=White,draw=Black,scale=0.8]
\tikzstyle{arrow_end}=[circle,fill=none,draw=none,scale=.1]
\tikzstyle{X}=[circle,fill=White,draw=none,label={[xshift=0 cm, yshift=-0.3 cm,Grey]X},scale=.1]
\tikzstyle{Z}=[circle,fill=White,draw=none,label={[xshift=0 cm, yshift=-0.3 cm,Grey]Z},scale=.1]
\tikzstyle{a}=[circle,fill=none,draw=none,scale=.1,label={[xshift=0 cm, yshift=-0.5 cm]a)}]
\tikzstyle{b}=[circle,fill=none,draw=none,scale=.1,label={[xshift=0 cm, yshift=-0.5 cm]b)}]
\tikzstyle{c}=[circle,fill=none,draw=none,scale=.1,label={[xshift=0 cm, yshift=-0.5 cm]c)}]
\tikzstyle{eq}=[circle,fill=none,draw=none,scale=2,label={[xshift=0 cm, yshift=-0.5 cm]=}]
\tikzstyle{cir}=[circle,fill=none,draw=Blue,dashed,scale=4,line width=2 pt]
\definecolor{tempcolor}{rgb}{.9,.7,0}
\tikzstyle{cirX}=[circle,fill=none,draw=tempcolor,scale=4,line width=2 pt]
\tikzstyle{cirZ}=[circle,fill=none,draw=DeepSkyBlue,scale=4,line width=2 pt]
\tikzstyle{dotdotdot}=[circle,fill=White,draw=none,label={[xshift=0 cm, yshift=-0.3 cm,Black]...},scale=1]
\tikzstyle{h1}=[circle,fill=White,draw=none,label={[xshift=0 cm, yshift=-0.3 cm,Grey]$h_1$},scale=.1]
\tikzstyle{h2}=[circle,fill=White,draw=none,label={[xshift=0 cm, yshift=-0.3 cm,Grey]$h_2$},scale=.1]
\tikzstyle{J}=[circle,fill=White,draw=none,label={[xshift=0 cm, yshift=-0.3 cm,Grey]$J$},scale=.1]

\tikzstyle{simple}=[-,draw=Black,line width=1.000]
\tikzstyle{arrow}=[-,draw=Black,postaction={decorate},decoration={markings,mark=at position .5 with {\arrow{>}}},line width=2.000]
\tikzstyle{tick}=[-,draw=Black,postaction={decorate},decoration={markings,mark=at position .5 with {\draw (0,-0.1) -- (0,0.1);}},line width=2.000]
\tikzstyle{conj_prop}=[-,draw=Black,line width=1.00]
\tikzstyle{CNOT}=[-,draw=Red,line width=1.00]
\tikzstyle{propagation}=[-,draw=Magenta,postaction={decorate},decoration={markings,mark=at position .5 with {\arrow{>}}},line width=1.00]
\tikzstyle{big_arrow}=[-latex,draw=Black,line width=3.000]
\tikzstyle{thin_arrow}=[->,draw=Grey,dash pattern=on \pgflinewidth off 2pt,line width=1]
\tikzstyle{undetected}=[dashed,postaction={decorate},decoration={markings,mark=at position .5 with {\arrow{>}}},draw=Grey,line width=1.5]
\tikzstyle{arrow_grey}=[->,draw=Grey,dash pattern=on \pgflinewidth off 2pt,line width=1]
\tikzstyle{added_edge}=[-,draw=Blue,line width=1.5]
\tikzstyle{cancelled_edge}=[-,draw=none,dashed,line width=1.5]
\tikzstyle{indirect_prop}=[-,draw=Blue,line width=1.5]
\definecolor{tempcolor}{rgb}{.9,.7,0}
\tikzstyle{added_S1}=[-,draw=tempcolor,line width=1.5]
\tikzstyle{added_S2}=[-,draw=DeepSkyBlue,line width=1.5]

%% file: tikzit/data_qubit.tikz
\begin{tikzpicture}
	\begin{pgfonlayer}{nodelayer}
		\node [style={data_qubit}] (0) at (0, -0) {};
	\end{pgfonlayer}
\end{tikzpicture}

%% file: tikzit/parity_qubit.tikz
\begin{tikzpicture}
	\begin{pgfonlayer}{nodelayer}
		\node [style={parity_check_qubit}] (0) at (0, -0) {};
	\end{pgfonlayer}
\end{tikzpicture}

%% file: tikzit/unmeasured_definition.tikz
\begin{tikzpicture}
	\begin{pgfonlayer}{nodelayer}
		\node [style={classical_data_bit}] (0) at (6.75, 7.5) {};
		\node [style={arrow_end}] (1) at (6, 7.5) {};
		\node [style={arrow_end}] (2) at (3, 7.5) {};
		\node [style={classical_data_phase}] (3) at (7.25, 7.5) {};
		\node [style={arrow_end}] (4) at (3, 7.5) {};
		\node [style={data_qubit}] (5) at (2, 7.5) {};
	\end{pgfonlayer}
	\begin{pgfonlayer}{edgelayer}
		\draw [style={big_arrow}] (2) to (1);
	\end{pgfonlayer}
\end{tikzpicture}

%% file: tikzit/measurement_definition.tikz
\begin{tikzpicture}
	\begin{pgfonlayer}{nodelayer}
		\node [style={arrow_end}] (0) at (3, -4) {};
		\node [style={classical_parity_check}] (1) at (7, -4) {};
		\node [style={parity_check_qubit}] (2) at (2, -4) {};
		\node [style={arrow_end}] (3) at (3, -4) {};
		\node [style={arrow_end}] (4) at (6, -4) {};
		\node [style={classical_data_phase}] (5) at (7.5, -4) {};
	\end{pgfonlayer}
	\begin{pgfonlayer}{edgelayer}
		\draw [style={big_arrow}] (3) to (4);
	\end{pgfonlayer}
\end{tikzpicture}

%% file: tikzit/measurement_definition_self.tikz
\begin{tikzpicture}
	\begin{pgfonlayer}{nodelayer}
		\node [style={arrow_end}] (0) at (3, -4) {};
		\node [style={classical_parity_check}] (1) at (6.75, -4) {};
		\node [style={parity_check_qubit}] (2) at (2, -4) {};
		\node [style={arrow_end}] (3) at (3, -4) {};
		\node [style={arrow_end}] (4) at (6, -4) {};
		\node [style={classical_data_phase}] (5) at (7.25, -4) {};
	\end{pgfonlayer}
	\begin{pgfonlayer}{edgelayer}
		\draw [style={big_arrow}] (3) to (4);
		\draw [style=propagation, in=135, out=45, loop] (2) to ();
		\draw [style=simple] (1) to (5);
	\end{pgfonlayer}
\end{tikzpicture}

%% file: tikzit/unmeasured_definition_Y.tikz
\begin{tikzpicture}
	\begin{pgfonlayer}{nodelayer}
		\node [style={classical_data_bit}] (0) at (6.75, 7.25) {};
		\node [style={arrow_end}] (1) at (6, 7.5) {};
		\node [style={arrow_end}] (2) at (3, 7.5) {};
		\node [style={classical_data_phase}] (3) at (7.25, 7.25) {};
		\node [style={arrow_end}] (4) at (3, 7.5) {};
		\node [style={data_qubit}] (5) at (2, 7.5) {};
		\node [style={classical_parity_checkY}] (6) at (7, 7.75) {};
	\end{pgfonlayer}
	\begin{pgfonlayer}{edgelayer}
		\draw [style={big_arrow}] (2) to (1);
		\draw [style=simple] (0) to (6);
		\draw [style=simple] (3) to (6);
	\end{pgfonlayer}
\end{tikzpicture}

%% file: tikzit/measurement_definition_Y.tikz
\begin{tikzpicture}
	\begin{pgfonlayer}{nodelayer}
		\node [style={arrow_end}] (0) at (3, -4) {};
		\node [style={classical_parity_check}] (1) at (7, -4) {};
		\node [style={parity_check_qubit}] (2) at (2, -4) {};
		\node [style={arrow_end}] (3) at (3, -4) {};
		\node [style={arrow_end}] (4) at (6, -4) {};
		\node [style={classical_data_phase}] (5) at (7.5, -4) {};
		\node [style={classical_parity_checkY}] (6) at (8, -4) {};
		\node [style={arrow_end}] (7) at (1.5, -4.5) {};
		\node [style={arrow_end}] (8) at (2, -4.5) {};
		\node [style={arrow_end}] (9) at (2.5, -4.5) {};
		\node [style={arrow_end}] (10) at (7.75, -4.75) {};
		\node [style={arrow_end}] (11) at (7.5, -4.75) {};
		\node [style={arrow_end}] (12) at (7.25, -4.75) {};
	\end{pgfonlayer}
	\begin{pgfonlayer}{edgelayer}
		\draw [style={big_arrow}] (3) to (4);
		\draw [style=simple] (7) to (2);
		\draw [style=simple] (8) to (2);
		\draw [style=simple] (9) to (2);
		\draw [style=simple] (6) to (10);
		\draw [style=simple] (6) to (11);
		\draw [style=simple] (6) to (12);
		\draw [style=simple] (6) to (5);
		\draw [style=simple] (1) to (12);
		\draw [style=simple] (1) to (11);
		\draw [style=simple] (1) to (10);
	\end{pgfonlayer}
\end{tikzpicture}

%% file: tikzit/measurement_definition_Yself.tikz
\begin{tikzpicture}
	\begin{pgfonlayer}{nodelayer}
		\node [style={arrow_end}] (0) at (3, -4) {};
		\node [style={classical_parity_check}] (1) at (7, -4) {};
		\node [style={parity_check_qubit}] (2) at (2, -4) {};
		\node [style={arrow_end}] (3) at (3, -4) {};
		\node [style={arrow_end}] (4) at (6, -4) {};
		\node [style={classical_data_phase}] (5) at (7.5, -4) {};
		\node [style={classical_parity_checkY}] (6) at (8, -4) {};
		\node [style={arrow_end}] (7) at (1.5, -4.5) {};
		\node [style={arrow_end}] (8) at (2, -4.5) {};
		\node [style={arrow_end}] (9) at (2.5, -4.5) {};
		\node [style={arrow_end}] (10) at (7.75, -4.75) {};
		\node [style={arrow_end}] (11) at (7.5, -4.75) {};
		\node [style={arrow_end}] (12) at (7.25, -4.75) {};
	\end{pgfonlayer}
	\begin{pgfonlayer}{edgelayer}
		\draw [style={big_arrow}] (3) to (4);
		\draw [style=simple] (7) to (2);
		\draw [style=simple] (8) to (2);
		\draw [style=simple] (9) to (2);
		\draw [style=simple] (6) to (10);
		\draw [style=simple] (6) to (11);
		\draw [style=simple] (6) to (12);
		\draw [style=simple] (1) to (12);
		\draw [style=simple] (1) to (11);
		\draw [style=simple] (1) to (10);
		\draw [style=propagation, in=135, out=45, loop] (2) to ();
		\draw [style=simple] (5) to (1);
	\end{pgfonlayer}
\end{tikzpicture}

%% file: tikzit/over_time.tikz
\begin{tikzpicture}
	\begin{pgfonlayer}{nodelayer}
		\node [style={classical_data_bit}] (0) at (1.5, 4.75) {};
		\node [style={classical_data_phase}] (1) at (1.5, 5.75) {};
		\node [style={classical_parity_check}] (2) at (1.5, 5.25) {};
		\node [style={classical_data_bit}] (5) at (3, 4.75) {};
		\node [style={classical_data_phase}] (6) at (3, 5.75) {};
		\node [style={classical_parity_check}] (7) at (2.25, 5.25) {};
		\node [style={classical_parity_check}] (8) at (2.25, 6) {};
		\node [style={classical_parity_check}] (9) at (2.25, 4.5) {};
		\node [style={classical_data_phase}] (10) at (4.5, 5.75) {};
		\node [style={classical_data_bit}] (11) at (4.5, 4.75) {};
		\node [style={classical_parity_check}] (12) at (3.75, 6) {};
		\node [style={classical_parity_check}] (13) at (3.75, 5.25) {};
		\node [style={classical_parity_check}] (14) at (3.75, 4.5) {};
		\node [style=h2] (15) at (1, 5.75) {};
		\node [style=h1] (16) at (1, 4.75) {};
		\node [style=J] (17) at (1, 5.25) {};
		\node [style=J] (18) at (2.75, 5.25) {};
		\node [style=J] (19) at (3.25, 5.25) {};
		\node [style={arrow_end}] (20) at (1.5, 4) {};
		\node [style={arrow_end}] (21) at (1.75, 4) {};
		\node [style={arrow_end}] (22) at (1.25, 4) {};
		\node [style={arrow_end}] (23) at (1.25, 6.5) {};
		\node [style={arrow_end}] (24) at (1.5, 6.5) {};
		\node [style={arrow_end}] (25) at (1.75, 6.5) {};
		\node [style={arrow_end}] (26) at (3.25, 6.5) {};
		\node [style={arrow_end}] (27) at (2.75, 6.5) {};
		\node [style={arrow_end}] (28) at (3, 6.5) {};
		\node [style={arrow_end}] (29) at (4.75, 6.5) {};
		\node [style={arrow_end}] (30) at (4.25, 6.5) {};
		\node [style={arrow_end}] (31) at (4.5, 6.5) {};
		\node [style={arrow_end}] (32) at (3.25, 4) {};
		\node [style={arrow_end}] (33) at (2.75, 4) {};
		\node [style={arrow_end}] (34) at (3, 4) {};
		\node [style={arrow_end}] (35) at (4.75, 4) {};
		\node [style={arrow_end}] (36) at (4.25, 4) {};
		\node [style={arrow_end}] (37) at (4.5, 4) {};
		\node [style=dotdotdot] (38) at (5.25, 5.25) {};
		\node [style=h2] (39) at (2.25, 6.5) {};
		\node [style=h2] (40) at (3.75, 6.5) {};
		\node [style=h1] (41) at (2.25, 4) {};
		\node [style=h1] (42) at (3.75, 4) {};
	\end{pgfonlayer}
	\begin{pgfonlayer}{edgelayer}
		\draw [style=simple] (0) to (2);
		\draw [style=simple] (1) to (2);
		\draw [style=simple] (0) to (7);
		\draw [style=simple] (1) to (7);
		\draw [style=simple] (7) to (6);
		\draw [style=simple] (7) to (5);
		\draw [style=simple] (0) to (9);
		\draw [style=simple] (9) to (5);
		\draw [style=simple] (1) to (8);
		\draw [style=simple] (8) to (6);
		\draw [style=simple] (13) to (10);
		\draw [style=simple] (13) to (11);
		\draw [style=simple] (14) to (11);
		\draw [style=simple] (12) to (10);
		\draw [style=simple] (5) to (14);
		\draw [style=simple] (6) to (12);
		\draw [style=simple] (6) to (13);
		\draw [style=simple] (5) to (13);
		\draw [style=simple] (22) to (0);
		\draw [style=simple] (0) to (20);
		\draw [style=simple] (0) to (21);
		\draw [style=simple] (1) to (23);
		\draw [style=simple] (1) to (24);
		\draw [style=simple] (1) to (25);
		\draw [style=simple] (6) to (27);
		\draw [style=simple] (6) to (28);
		\draw [style=simple] (6) to (26);
		\draw [style=simple] (10) to (30);
		\draw [style=simple] (10) to (30);
		\draw [style=simple] (10) to (31);
		\draw [style=simple] (10) to (29);
		\draw [style=simple] (11) to (36);
		\draw [style=simple] (11) to (37);
		\draw [style=simple] (11) to (35);
		\draw [style=simple] (5) to (33);
		\draw [style=simple] (5) to (34);
		\draw [style=simple] (5) to (32);
	\end{pgfonlayer}
\end{tikzpicture}